\documentclass[%
 reprint,
 amsmath,amssymb,
 aps,
]{revtex4-2}

\usepackage[final]{graphicx} 

\begin{document}

\preprint{APS/123-QED}

\title{Proximity driven photon-tunneling in chiral quantum hybrid systems}

\author{Aryan Pratap Srivastava}
\thanks{These authors contributed equally to this work.}
\affiliation{Nano‐Magnetism and Quantum Technology Lab, Department of Physics, Indian Institute of Technology (Banaras Hindu University), Varanasi 221005, India}

\author{Moulik Deviprasad Ketkar}
\thanks{These authors contributed equally to this work.}
\affiliation{Nano‐Magnetism and Quantum Technology Lab, Department of Physics, Indian Institute of Technology (Banaras Hindu University), Varanasi 221005, India}

\author{Kuldeep Kumar Shrivastava}
\affiliation{Nano‐Magnetism and Quantum Technology Lab, Department of Physics, Indian Institute of Technology (Banaras Hindu University), Varanasi 221005, India}

\author{Abhishek Maurya}
\affiliation{Nano‐Magnetism and Quantum Technology Lab, Department of Physics, Indian Institute of Technology (Banaras Hindu University), Varanasi 221005, India}

\author{Biswanath Bhoi}
\affiliation{Nano‐Magnetism and Quantum Technology Lab, Department of Physics, Indian Institute of Technology (Banaras Hindu University), Varanasi 221005, India}

\author{Rajeev Singh}
\email{rajeevs.phy@itbhu.ac.in}
\affiliation{Nano‐Magnetism and Quantum Technology Lab, Department of Physics, Indian Institute of Technology (Banaras Hindu University), Varanasi 221005, India}

\date{\today}

\begin{abstract}
In chiral quantum hybrid systems, resonator geometry and spacing critically govern photon-mediated interactions and enable engineered light–matter coupling. We experimentally explore proximity-driven photon tunneling in a pair of coupled, inverted circular split-ring microwave resonators with discrete chiral orientations ($\phi = 0^\circ$, $90^\circ$, $180^\circ$, $270^\circ$). These resonators, coherently excited via a shared photon guide, serve as a platform for studying directional and symmetry-dependent coupling. By systematically varying inter-resonator spacing, we observe pronounced modulation in transmission spectra, with coupling strength strongly influenced by both chirality and proximity. Microwave spectroscopy on fabricated devices reveals distinct hybridization behavior, including mode splitting, interference, and the emergence of dark states. These findings are further supported by full-wave electromagnetic simulations. To interpret the observed phenomena, we develop a circuit quantum electrodynamics (cQED) model that accurately captures tunneling behavior, coupling sign reversal, and spectral features. This comprehensive experimental, numerical, and theoretical investigation advances the understanding of phase- and geometry-controlled interactions in chiral hybrid systems. The system behaves as a classical analogue of a chiral quantum hybrid platform, where photon tunneling, mode hybridization, and chirality-dependent phase interference follow the same structure as two quantized harmonic oscillators. Our use of a cQED formalism reflects the quantum-consistent nature of these interactions, even though the experimental excitation itself is classical. The demonstrated control over photon tunneling through structural and excitation parameters paves the way for reconfigurable photonic devices with applications in quantum communication, chiral biosensing, and polarization-selective signal processing.
\end{abstract}

\maketitle

\section{Introduction}

In the rapidly evolving field of quantum photonics, controlling light--matter interactions at the quantum level is crucial for developing next-generation quantum communication, information processing, and sensing devices. Chiral quantum hybrid systems---where symmetry-breaking mechanisms enable direction-dependent coupling between photons and resonators or quantum emitters~\cite{Zhao2018, Galiffi2024, Scheucher2016, Gao2023}---offer a promising path. These systems exploit spin--momentum locking and structural asymmetry to realize selective emission and controlled photon transport~\cite{Nagaosa2024, Chen2020, Krasnok2022}, laying the foundation for compact, on-chip quantum technologies. A defining feature of these systems is their sensitivity to the geometry and symmetry of resonator elements. Specifically, the chirality---determined by handedness or angular orientation of structures like split rings---strongly affects polarization, directionality, and coupling strength~\cite{Lininger2023}. In photonic structures supporting circularly polarized modes, the spatial configuration of chiral elements enables direct control over modal overlap and field orientation, which in turn governs the efficiency and characteristics of inter-resonator coupling~\cite{Baranov2017}. Thus, engineering geometric parameters such as angular orientation, spacing, and symmetry can modulate coupling strength, spectral response, and coherent mode formation~\cite{Khosravi2019}. While classical and semiclassical studies have shown that structural chirality in split-ring resonators induces optical activity, mode splitting, and polarization-selective transmission~\cite{Kass2024}, a complete quantum-mechanical understanding of photon tunneling between closely spaced chiral resonators remains limited. Classical models such as coupled-mode theory fail to fully capture the quantum features of photon exchange at subwavelength separations, particularly those arising from asymmetric phase relations due to chirality. Although our experiments use classical, high-photon-number microwave excitation, the coupled IC-SRRs behave as two harmonic oscillators whose interaction is formally equivalent to that of a quantum Hamiltonian. In this linear regime, the dynamics—normal-mode splitting, hybridization, and chirality-dependent phase interference—match the predictions of a cQED framework, which compactly captures geometry-induced phase factors and coupling-sign reversal. Thus, the system serves as a classical analogue of a reciprocal quantum device, allowing chirality-dependent spectral features to be interpreted through a quantum-consistent interference picture while remaining fully compatible with classical microwave excitation. \\
In this work, we present a detailed study of proximity-driven photon tunneling in a minimal chiral quantum hybrid system composed of two inverted circular split-ring resonators (IC-SRRs), each oriented at one of four discrete chiral angles ($\phi = 0^\circ, 90^\circ, 180^\circ, 270^\circ$). Coupled via a common photon guide, these resonators support coherent excitation and near-field interactions. Using full-wave electromagnetic simulations, we examine how inter-resonator spacing and relative angular orientation influence coupling strength, mode hybridization, and spectral response. The simulations reveal phenomena such as mode splitting and chirality-dependent spectral shifts, which are validated through microwave spectroscopy measurements on fabricated structures. To interpret these results, we develop a circuit quantum electrodynamics (cQED) model that treats each resonator as a quantized harmonic oscillator, with coupling strengths explicitly dependent on both inter-resonator distance and relative chirality. The model accurately predicts resonance splitting, polarization properties, and tunneling dynamics. The agreement across simulation, theory, and experiment supports the photon-tunneling picture and establishes chirality as a tunable control parameter for manipulating mode interactions. The paper is organized as follows: Section~II describes the resonator design and simulation/fabrication procedures. Section~III presents the experimental transmission measurements. Section~IV introduces the quantum theoretical model, and Section~V discusses implications and potential applications in quantum technologies.

\section{Numerical Simulation of the Hybrid Device}

\subsection{Design and Layout of the Hybrid Device}
The device comprises a three-layered planar structure~\cite{trovatello2024tunable}, shown in Fig.~1(a). The top layer is a copper microstrip line (SnPb-coated) for microwave excitation and readout. The middle FR-4 dielectric~\cite{soci2023roadmap} ensures minimal attenuation of evanescent fields, while the bottom copper ground plane hosts two inverted circular split-ring resonators (IC-SRRs) that support localized quantum-like modes critical for near-field coupling and photon tunneling [Fig.~1(b)]. To investigate chirality effects, one IC-SRR is fixed at angle $\phi_1$, and the other is rotated to $\phi_2 = 0^\circ, 90^\circ, 180^\circ$, or $270^\circ$, defining four configurations: P, Q, R, and S [Fig.~1(c)]. Simulations and measurements are performed independently for each to study how the relative chirality $\Delta\phi = \phi_2 - \phi_1$ influences coupling and spectra. As the spacing $d$ decreases, evanescent field overlap increases, enhancing photon tunneling and promoting mode hybridization. This behavior is governed jointly by $d$ and the chirality-induced phase shift $\Delta\phi$, and is modeled within a circuit QED framework where each IC-SRR is treated as a quantized harmonic oscillator. The coupling coefficient $\Delta_{AB}$ is expressed as: \begin{equation}
\Delta_{AB}(d,\Delta\phi)
=
\Theta(\Delta\phi)
\exp\!\left(-\frac{d}{d_0}\right).
\end{equation} where $\Theta(\Delta\phi)$ captures chirality-dependent phase modulation and ${d_0}$ can be understood as the characteristic length within the system without loss of any generality, ensuring consistency in the interpretation of spatial parameters such as resonator separation. This model explains the observed spectral features: maximal mode splitting at $\Delta\phi = 180^\circ$ due to destructive interference, asymmetric evolution at $90^\circ$ and $270^\circ$, and minimal hybridization at $0^\circ$ from constructive alignment. 
\begin{figure} 
\includegraphics[width=\columnwidth]{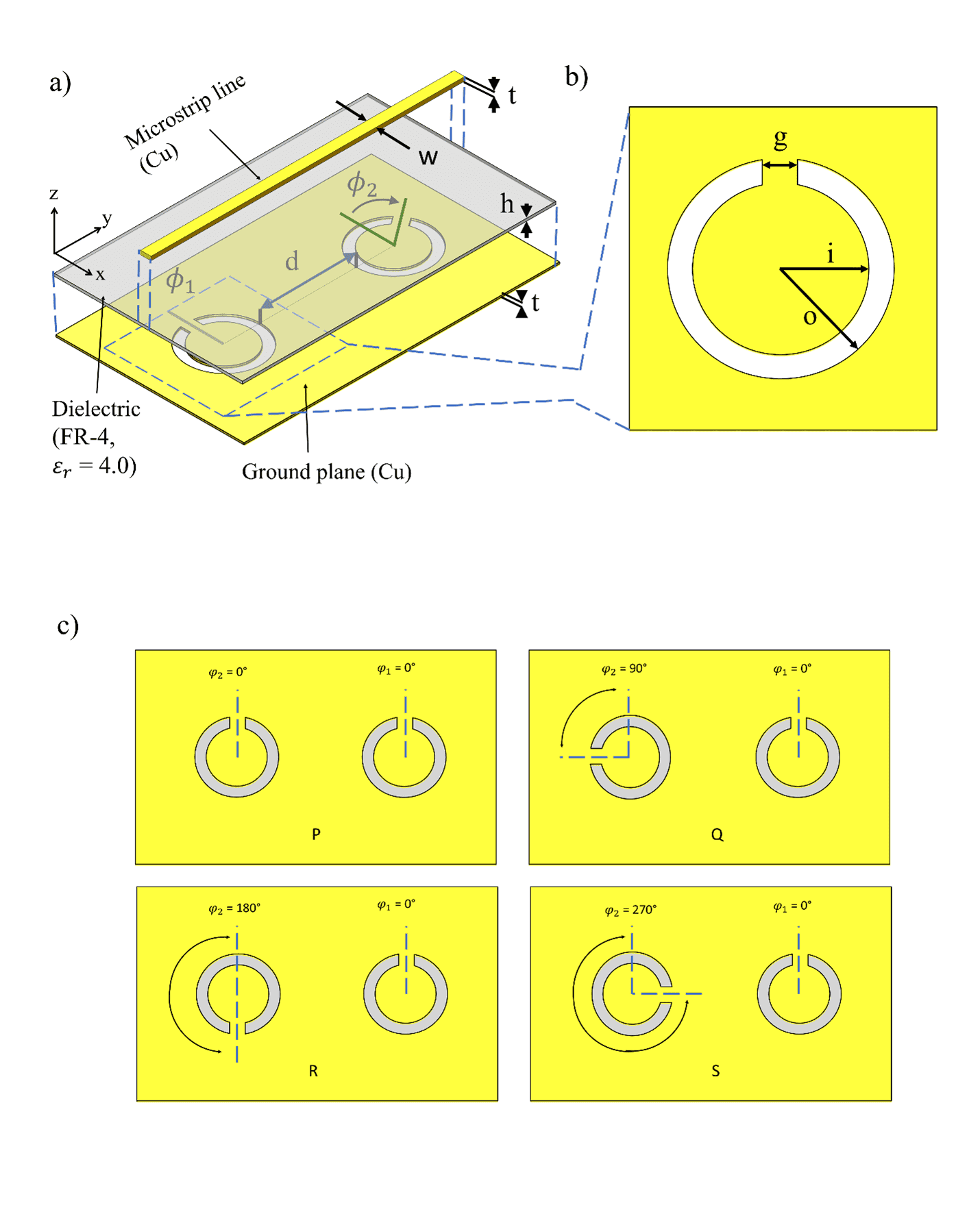} \caption{(a) Schematic illustrations of the sample structure comprising a microstrip line (dark yellow), an FR-4 dielectric substrate (gray), and two identical ISRRs patterned on the ground plane (also dark yellow). The split gaps of the ISRRs are oriented at angles $\phi_1$ and $\phi_2$ relative to the $-x$ axis. The inter-resonator spacing $d$ varies from 0 to 10 mm across 100 configurations. The substrate thickness $h$ is 1.6 mm, the copper thickness $t$ for both the microstrip line and ground plane is 0.035 mm, and the microstrip line width $w$ is 3.137 mm. (b) Enlarged view of a single ISRR showing key dimensions: the split gap width $g$ = 0.15 mm, the inner radius $i$ = 2.0 mm, and the outer radius $o$ = 3.1 mm. (c) Four distinct chiral configurations of the IC-SRR pair representing different angular orientations with respect to the $-x$ axis.} \label{fig1} \end{figure}

\subsection{Simulation Details and Analysis}
Numerical simulations were performed using CST Microwave Studio to investigate the coupling behavior of interacting IC-SRRs across different chiral configurations. For each case, the center-to-center separation distance $d$ (excluding the resonator radii) was varied from 0 to 10 mm, sampled over 100 points to capture fine variations in coupling strength and modal evolution. Four configurations were studied, corresponding to relative chiral angles $\Delta\phi = \phi_2 - \phi_1 = 0^\circ, 90^\circ, 180^\circ$, and $270^\circ$, labeled P, Q, R, and S respectively. As a representative example, Fig.~2(a) shows the transmission response for $\Delta\phi = 180^\circ$, revealing two distinct resonant peaks (green and orange) at small $d$ due to strong photon-mediated interactions and hybridized mode formation. As $d$ increases, dipolar field overlap diminishes, leading to weaker coupling and eventual convergence of the split resonances. This behavior is visualized in the 2D transmission colormap in Fig.~2(b), with vertical dashed lines indicating the spectra shown in Fig.~2(a), which displays only a discrete set of individual transmission traces extracted from the full simulation dataset at selected values of $d$, whereas Fig.~2(b) represents the continuous CST-generated frequency–distance colormap. Distinct trends emerge across the other configurations, shown in Figs.~2(c)–2(e). For $\Delta\phi = 0^\circ$, identical dipole alignment results in no frequency splitting, though mode intensity increases with $d$. At $\Delta\phi = 90^\circ$, weak splitting is observed at small $d$, along with a subtle anti-crossing feature at intermediate distances—suggesting interaction with a nearby third mode. In the $\Delta\phi = 270^\circ$ case, clear splitting is again absent, but slight peak broadening and amplitude enhancement occur at small $d$. In all configurations, a diffuse high-frequency mode appears at small $d$, interacting more strongly with $\Delta\phi = 90^\circ$ and contributing to the observed anti-crossing. This third spectral branch likely arises from weak coupling to parasitic or higher-order modes, introducing additional complexity into the coupling landscape. 
\begin{figure}[t] 
\includegraphics[width=\columnwidth]{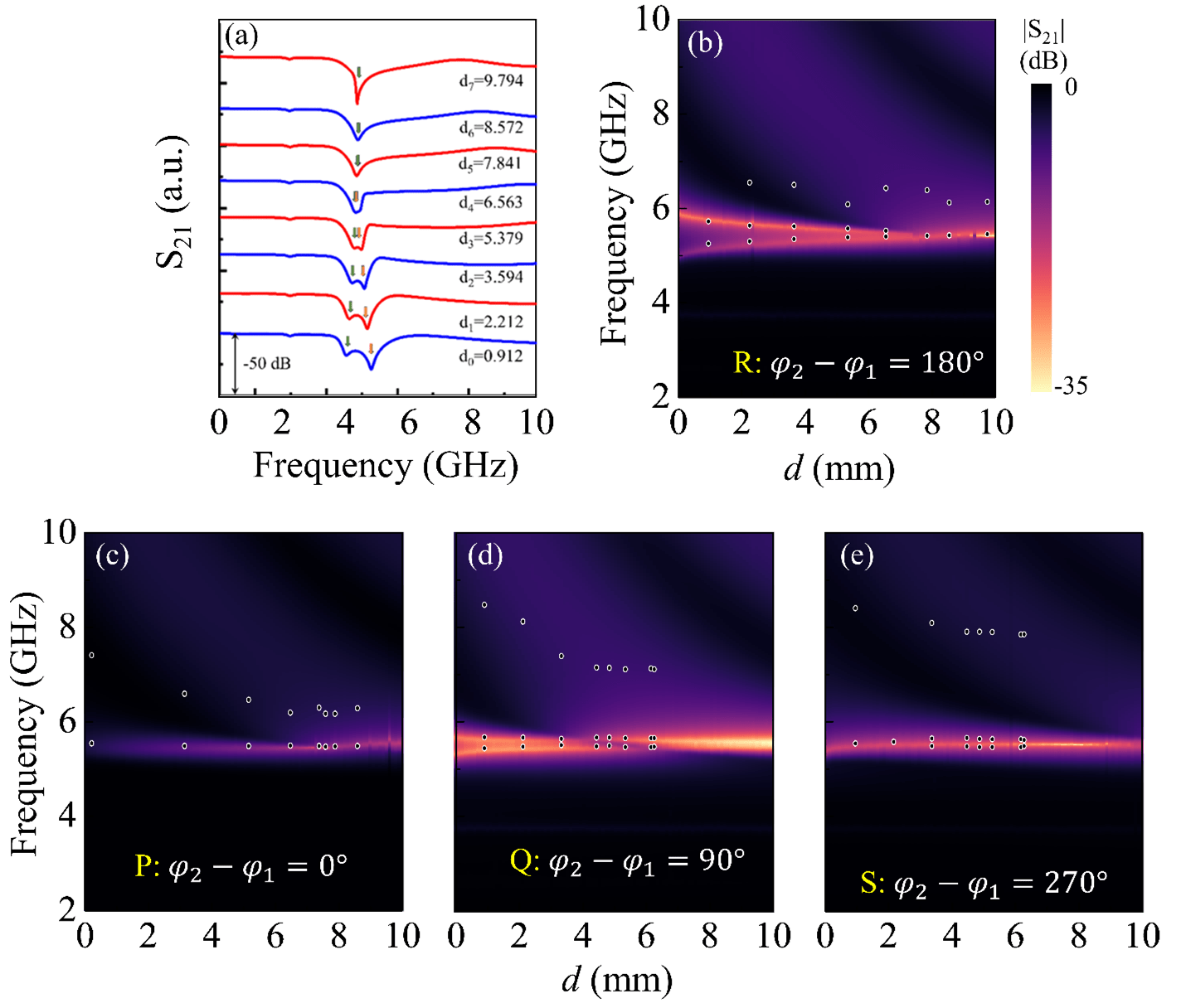} \caption{(a) Transmission spectra $|S_{21}|$ as a function of microwave frequency for varying inter-resonator spacing $'d'$ in configuration $R$ ($\phi_2-\phi_1 = 180^\circ$), illustrating the progressive merging of the split resonant peaks (highlighted by green and orange arrows). (b) Color map of $|S_{21}|$ power in the frequency–distance ( $f – d$) plane for configuration $R$, with red and blue dashed lines corresponding to the spectral traces shown in (a). (c–e) $|S_{21}|$ power maps in the $f – d$ plane for configurations: (c) $P$ ($\phi_2-\phi_1 = 0^\circ$), (d) $Q$ ($\phi_2-\phi_1 = 90^\circ$), (e) $S$ ($\phi_2-\phi_1 = 270^\circ$), respectively. The open circle symbols overlaid on the simulated $|S_{21}|$ color maps indicate the resonance frequencies extracted from the experimentally measured transmission spectra for selected samples corresponding to each chiral configuration of the hybrid samples.} 
\label{fig2} 
\end{figure}

\section{Experimental Details}

To experimentally investigate photon tunneling in coupled IC-SRRs across different chiral configurations, eight discrete inter-resonator spacings $d$ were selected for each of four relative orientations $\Delta\phi = \phi_2 - \phi_1 = 0^\circ, 90^\circ, 180^\circ$, and $270^\circ$, resulting in 32 distinct hybrid designs. Dimensional and design details of all fabricated samples are provided in TABLE 1 of the Supplementary Information (Section S1). The devices were fabricated using standard photolithography and PCB processing, with the ground-plane metallic regions precisely etched to form the IC-SRR patterns. Prior to measurements, the ends of each sample were filed with sandpaper to remove the SnPb coating, enhancing electrical contact with the connectors. Identical 1.6~mm female SMA connectors were then soldered to both ends of the microstrip line, serving as input and output ports for signal excitation. 

\begin{figure}[h]
    \centering
    \includegraphics[width=\columnwidth]{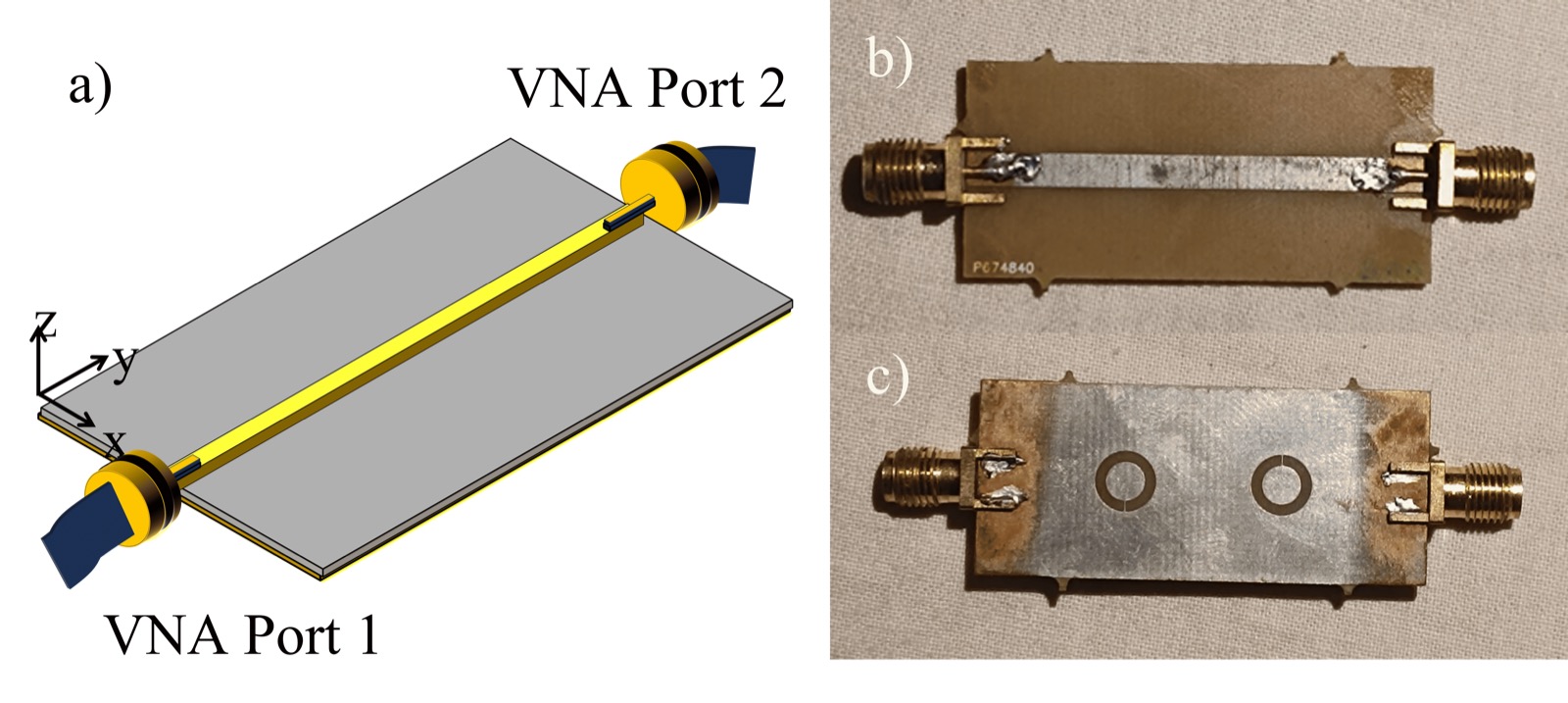}
    \caption{(a) Schematic diagram of the experimental setup used for transmission measurements.
    (b,c) Photographic images of a representative fabricated sample showing the front and back sides, respectively.}
    \label{fig:exp_setup}
\end{figure}

Figure~3(a) illustrates the experimental setup used to measure the transmission coefficient $|S_{21}|$, in which the input and output ends of the microstrip feed line are connected to a vector network analyzer (Rohde \& Schwarz ZNLE18). Figures~3(b) and 3(c) show optical micrographs of a representative fabricated sample with $\Delta\phi = 0^\circ$ and $d = 0.4$~mm, highlighting the structural layout on both sides of the PCB. Specifically, Fig.~3(b) displays the top layer containing only the microstrip excitation line, whereas Fig.~3(c) shows the bottom ground-plane layer on which the two IC-SRRs are patterned. Measurements were performed at room temperature using a 1~mW input microwave signal. Mode positions were identified by locating peak points in $S_{21}$ vs.\ frequency plots for each configuration and spacing. Experimentally observed resonance frequencies are overlaid as black circles on the simulated $|S_{21}|$ colormaps in Figs.~2(b)–2(e). The measurements showed excellent qualitative agreement with simulations, with only minor discrepancies in peak positions and bandwidths, attributed to impedance mismatches at coaxial-to-microstrip transitions and fabrication tolerances. We describe the interaction between the IC-SRRs using a quantum photon-tunneling framework. Unlike classical dipole models, our approach treats each resonator as a quantized harmonic oscillator, with coupling mediated by coherent photon exchange through near-field interactions. This tight-binding-like model captures the formation of hybridized modes with geometry-dependent frequency shifts. Specifically, it reveals how inter-resonator spacing and relative orientation---which define the system's geometric chirality---control the tunneling amplitude and resulting spectral features, in agreement with both simulation and experiment.

\section{Theoretical Formulation}

Several classical and semiclassical frameworks, such as coupled-mode theory and full-wave Maxwell simulations~\cite{wu2024exact,hagness1998fdtd}, have been extensively used to model energy exchange in coupled resonators through evanescent field overlap. These methods, particularly those based on classical dipole interactions, offer valuable physical intuition and can successfully reproduce many features of resonator coupling, including mode splitting and interference. While a classical explanation for the observations presented in this work is certainly possible, we adopt a quantum mechanical perspective—motivated by the inherently quantized nature of the resonators and the coherence of the excitation process. Within this framework, we employ a circuit quantum electrodynamics (cQED) formalism that enables a compact and physically transparent description of photon tunneling, mode hybridization, and chirality-dependent coupling. This approach captures the role of geometry-induced phase factors and facilitates an analytically tractable treatment of the system's dynamics, while remaining consistent with classical observations in the appropriate limits. We first construct a quantum framework that captures the formation of the two primary resonance branches, corresponding to coherent photon tunneling and hybridized fundamental modes. This analysis includes the influence of relative chirality and proximity on the tunneling amplitude and mode symmetry. We then extend the model to incorporate a third, more diffuse spectral branch, arising from incoherent processes such as mode interference, loss-induced asymmetries, and partial decoherence at intermediate coupling regimes. The extended model thus provides a unified quantum description of all observed spectral features.

\subsection{Two-Mode Analysis}
To understand photon behavior in chiral microwave cavity modes, we begin with a general theoretical model for two identical resonating modes, $A$ and $B$, each with resonant frequency $\omega_r$. Coupled through a coefficient $\Delta_{AB}$, the system Hamiltonian is

\begin{equation}
\frac{H}{\hbar} = (\omega_r - i\alpha)\hat{A}^\dagger\hat{A} + (\omega_r - i\beta)\hat{B}^\dagger\hat{B}
+ \Delta_{AB}(\hat{A}^\dagger\hat{B} + \hat{B}^\dagger\hat{A}),
\end{equation}

where $\hat{A}^\dagger$ and $\hat{B}^\dagger$ are the creation operators for the two resonators, and $\alpha$, $\beta$ represent their intrinsic dissipation rates~\cite{watanabe2024magnetic,zhang2024parity}.  
In matrix form,

\begin{equation}
\frac{H}{\hbar} =
\begin{bmatrix}
\hat{A}^\dagger & \hat{B}^\dagger
\end{bmatrix}
\begin{bmatrix}
\omega_r - i\alpha & \Delta_{AB} \\
\Delta_{AB} & \omega_r - i\beta
\end{bmatrix}
\begin{bmatrix}
\hat{A} \\
\hat{B}
\end{bmatrix}.
\end{equation}

Assuming $\hbar = 1$, the hybridized eigenfrequencies are

\begin{equation}
\epsilon_{\pm} = 
\frac{\tilde{\omega}_A + \tilde{\omega}_B}{2}
\pm \sqrt{\Delta_{AB}^2 +
\left( \frac{\tilde{\omega}_A - \tilde{\omega}_B}{2} \right)^2 },
\end{equation}

where $\tilde{\omega}_{A,B} = \omega_r - i(\alpha,\beta)$. Since the two resonators are nearly identical, this reduces to  
$\epsilon_{\pm} \approx \tilde{\omega}_r \pm \Delta_{AB}$.

\medskip
\noindent
\textbf{Chirality- and Distance-Dependent Coupling.}  
To model $\Delta_{AB}$, we draw inspiration from tight-binding models~\cite{ganeshan2015tightbinding, sakurai2021mqm,shrivastava2024emergence,yamada2014critical}, where tunneling decays exponentially with separation $d$. However, our system exhibits a distinct additional feature: the resonance splitting is strongly dependent on chirality even at zero separation. This indicates that the coupling contains an intrinsic angular component.

Moreover, because the device geometry repeats under a full rotation, the coupling must be periodic in the relative orientation $\Delta\phi$. Guided by these physical constraints, we constructed a dimensionless, $2\pi$-periodic angular prefactor
$\Theta(\Delta\phi)$ using trigonometric functions. Through inspection of the measured splitting, followed by iterative mathematical modeling and refinement, we obtained

\begin{equation}
\Theta(\Delta\phi) =
\frac{\Delta\phi}{2\pi}
\sin\!\left(\frac{\Delta \phi}{2}\right)
\left[
\sin\!\left(\frac{\Delta \phi}{2}\right)
+ \cos\!\left(\frac{\Delta \phi}{2}\right)
\right],
\end{equation} which captures the experimentally observed periodicity, sign changes, and symmetry constraints.

To ensure dimensional consistency, the angular dependence is kept entirely within $\Theta(\Delta\phi)$, while the distance dependence is modeled using a dimensionless exponential involving a characteristic decay length $d_0$:

\begin{equation}
\Delta_{AB}(d,\Delta\phi)
=
\Theta(\Delta\phi)
\exp\!\left(-\frac{d}{d_0}\right).
\end{equation}

In this factorized form, the exponential term describes evanescent photon tunneling as a function of separation, whereas $\Theta(\Delta\phi)$ encodes the chirality-induced phase relation. This provides a compact, physically transparent description of the coupling.

\medskip
\noindent
\textbf{Coupling to the Microwave Signal Line.}  
We now include coupling of both resonators to a microwave signal line (MSL), introducing operators $p_k^\dagger$ and $p_k$ for the external continuum of modes of frequency $\omega_k$. The full Hamiltonian becomes

\begin{equation}
\begin{split}
H &= (\omega_r - i\alpha)\hat{A}^\dagger \hat{A} + (\omega_r - i\beta)\hat{B}^\dagger \hat{B}
+ \Delta_{AB}(\hat{A}^\dagger \hat{B} + \hat{B}^\dagger \hat{A}) \\
&\quad + \int_{-\infty}^{\infty} \omega_k\, p_k^\dagger p_k \, dk \\
&\quad + \int_{-\infty}^{\infty} \left[
\lambda_A(\hat{A} p_k^\dagger + \hat{A}^\dagger p_k)
+ \lambda_B(\hat{B} p_k^\dagger + \hat{B}^\dagger p_k)
\right] dk,
\end{split}
\end{equation}

where $\lambda_A$ and $\lambda_B$ denote the coupling strengths between the resonators and the MSL. The rotating wave approximation has been applied to the interaction terms. A schematic of this system is shown in Fig.~4.

\begin{figure}
    \centering
    \includegraphics[width=\columnwidth]{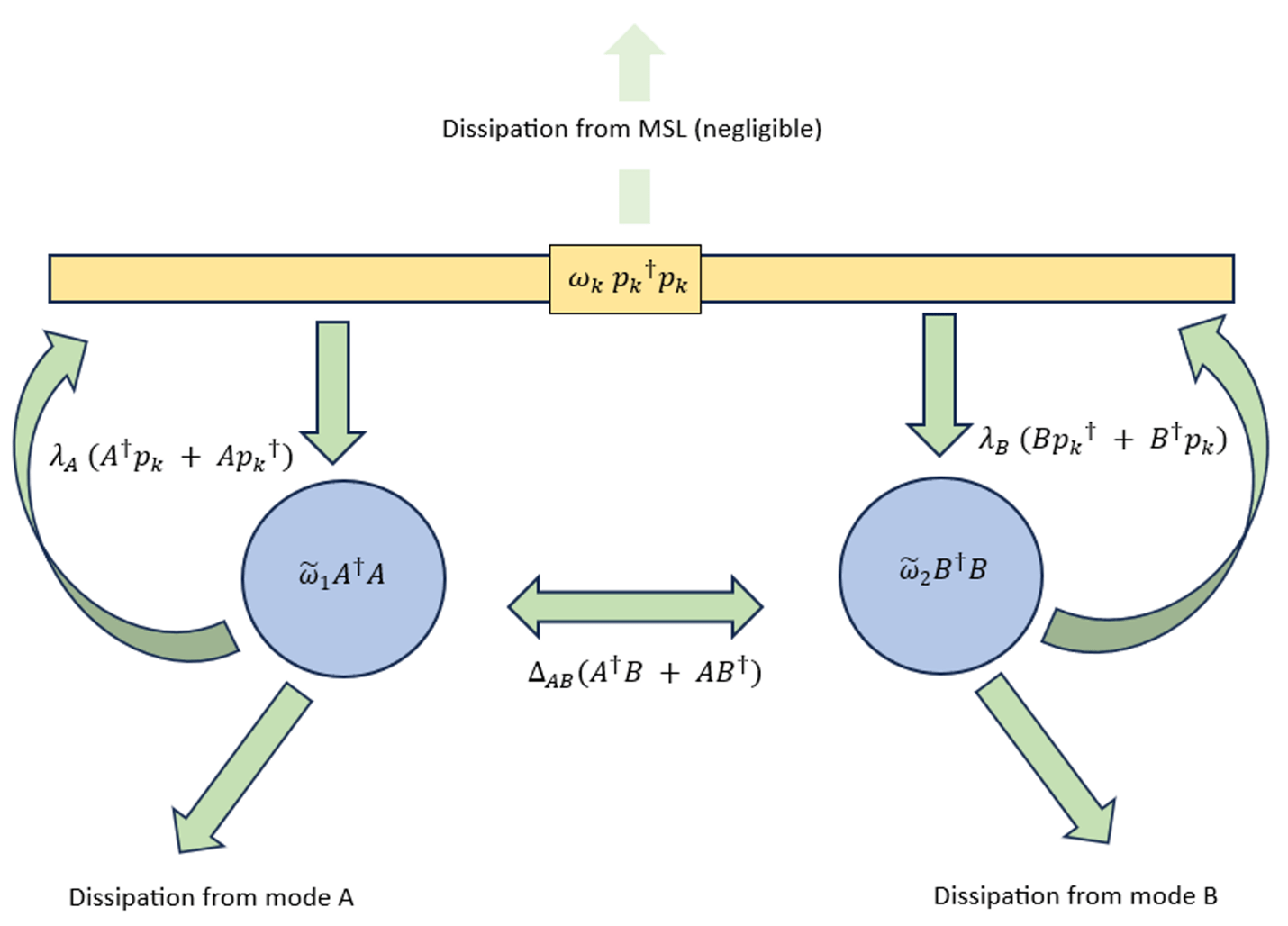}
    \caption{Schematic representation of the analytical quantum model illustrating two resonator modes coupled via a common signal line, along with relevant system parameters. All symbols are defined in detail within the main text.}
    \label{fig:analytical-cartoon}
\end{figure}

\medskip
\noindent
\textbf{Transmission Calculation.}  
Using the Heisenberg--Langevin formalism, the transmission coefficient is defined as

\begin{equation}
S_{21} = \frac{P_{\mathrm{out}}}{P_{\mathrm{in}}} - 1,
\end{equation}

with the input-output relation

\begin{equation}
\hat{P}_{\mathrm{out}}(\omega)
= \hat{P}_{\mathrm{in}}(\omega)
- 2i\bigl[ 
\sqrt{\beta_A}\,\hat{A}(\omega)
+ \sqrt{\beta_B}\,\hat{B}(\omega)
\bigr],
\end{equation}

providing a complete recipe for computing the analytical transmission spectrum following Ref.~\cite{shrivastava2024}.

\subsection{Three Mode Analysis}

The simulation plots reveal a diffuse third mode that intersects and interacts with the primary branches, introducing complex dynamics. This third mode can be interpreted as a quasi-bound leakage mode or indirect path interference, arising from chirality-dependent coupling. In particular, its contribution diminishes as the resonator separation increases. We extend our earlier model to account for the third mode in a very similar process as described above. (See Supplementary Material Section S4 ~\cite{sakurai2021mqm}for theoretical modeling of all the three modes). Figure 5, shows the theoretical color map of the transmission factor $S_{21}$ versus the frequency and separation of the resonators, helping us to verify the analytical modeling developed by us, which is in good agreement with the simulations for the four configurations considered. \begin{figure} \centering \includegraphics[width=\columnwidth]{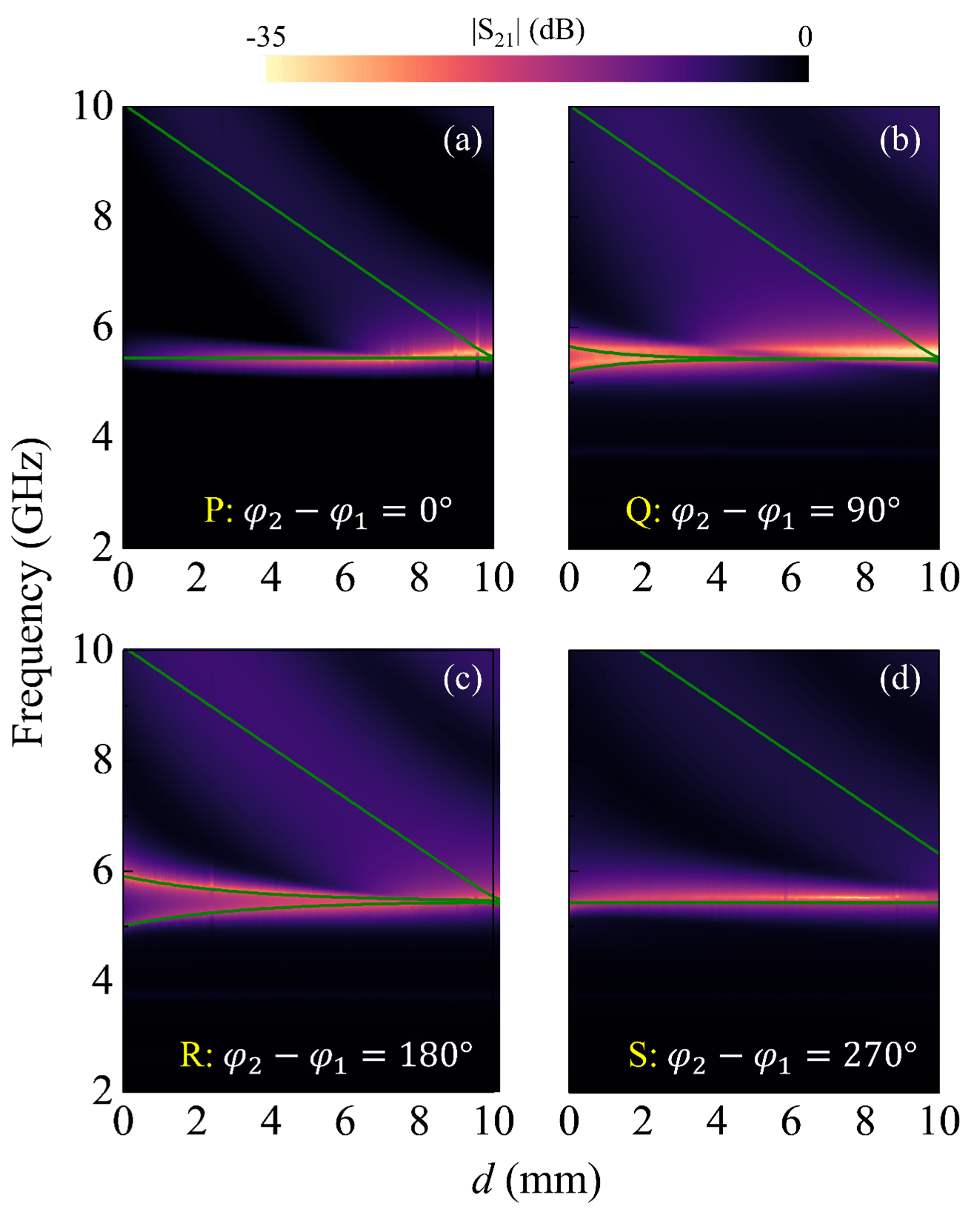} \caption{Theoretically calculated resonance frequencies (solid green lines) overlaid on the simulated $|S_{21}|$ color maps represent the for different chiral configurations of the hybrid samples: (a) $P$ ($\phi_2-\phi_1 = 0^\circ$) (b) $Q$ ($\phi_2-\phi_1 = 90^\circ$) (c) $R$ ($\phi_2-\phi_1 = 180^\circ$) and (d) $S$ ($\phi_2-\phi_1 = 270^\circ$), respectively.} \label{fig:analytical 3 mode} \end{figure}

\section{Results}

Our study demonstrates that photon tunneling in chiral IC-SRR systems is fundamentally governed by geometric chirality and excitation phase. Each IC-SRR functions as an LC resonator, where dipolar coupling is determined by the spatial separation and angular orientation~\cite{koch2010time}. At small separations, configuration P ($\phi = 0^\circ$) shows negligible hybridization due to identical dipole alignment, whereas configuration Q ($\phi = 90^\circ$) introduces a $\pi/2$ phase mismatch, yielding weak mode splitting. Configuration R ($\phi = 180^\circ$) supports strong antisymmetric hybridization with pronounced spectral splitting ~\cite{maksi1986symmetry}. In these cases, $\Delta_{AB} > 0$ reflects constructive tunneling, even as phase mismatches externally reduce observable coupling. Configuration S ($\phi = 270^\circ$), however, marks a transition to negative coupling, where $\Delta_{AB}$ becomes negative due to geometric phase inversion, suppressing symmetric hybridization and enabling antisymmetric (dark) modes~\cite{shrivastava2024}. This behavior is confirmed in Fig.~6 through analytical phase diagrams, which show how $\Delta_{AB}$ decays with distance and inverts as a function of chirality. 

\begin{figure}[h]
    \centering
    \includegraphics[width=\columnwidth]{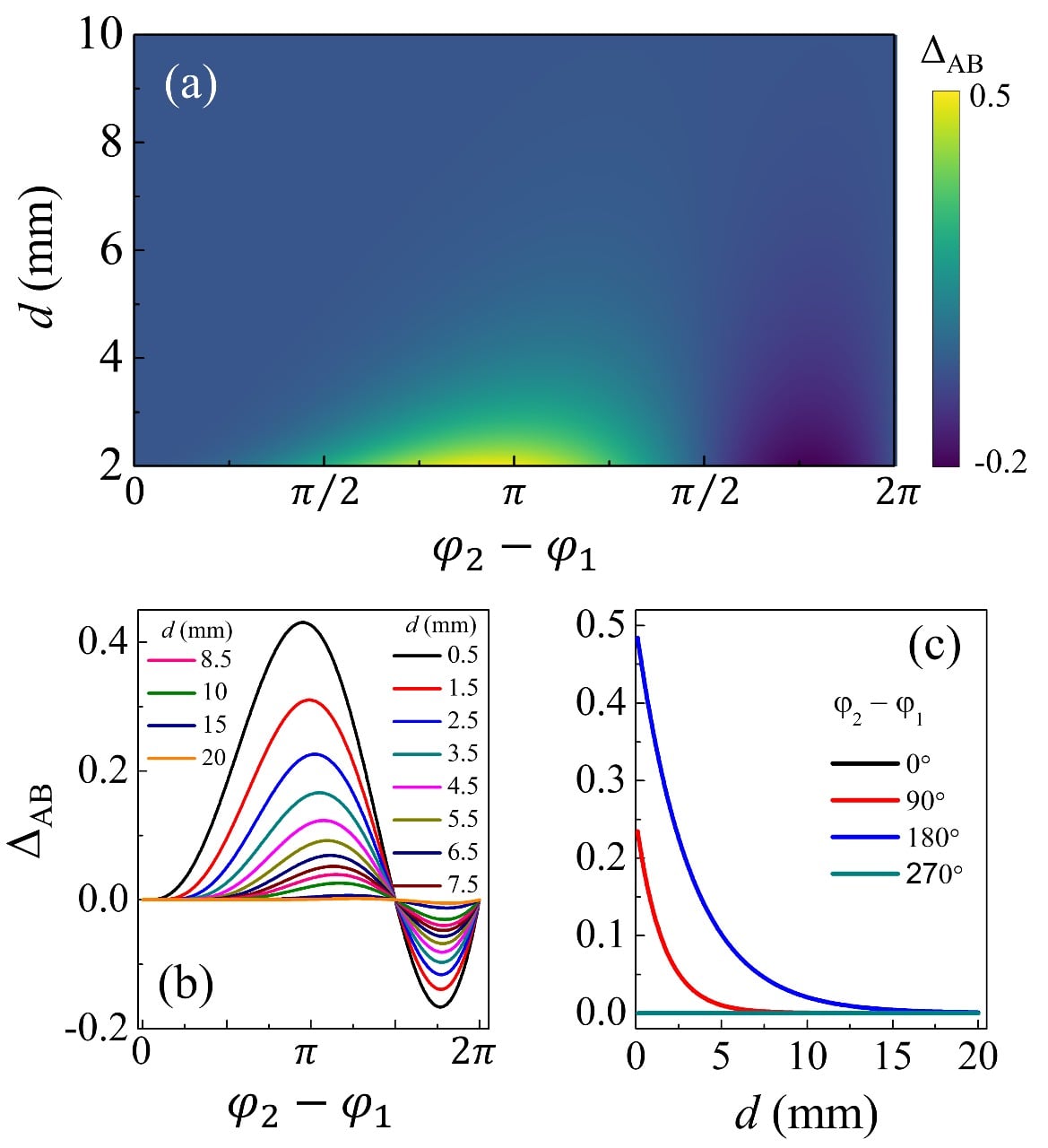}
    \caption{a) Analytically computed phase diagram of the coupling constant ($\Delta_{AB}$) plotted on the $d - (\phi_2-\phi_1)$ plane, where the color scale represents the magnitude of $\Delta_{AB}$ between the two hybrid modes. b) Variation of $\Delta_{AB}$ as a function of $\phi_2-\phi_1$ for selected values of inter-resonator spacing $d$. c) Variation of $\Delta_{AB}$ as a function of $d$ for different values of $\phi_2-\phi_1$.}
\label{fig:deltaAB_vs_phi_and_distance}

\end{figure}

Importantly, even with a fixed geometry ($\phi = 300^\circ$), we find that the excitation phase $\theta$ can dynamically modulate the coupling strength. At $\theta = 180^\circ$, hybridization partially persists due to incomplete destructive interference~\cite{liu2014tunneling}, whereas at $\theta = 270^\circ$, perfect phase cancellation leads to a dark state~\cite{Tang2024}. This dynamic switching—achieved purely by tuning $\theta$—demonstrates coherent control of light–matter interaction without structural alteration. Conceptually, the IC-SRR platform operates analogously to a Mach--Zehnder interferometer~\cite{nazarov1993quantum}, where photon interference arises from a competition between the tunneling phase (set by $\phi$) and excitation phase (set by $\theta$), inferring that the IC-SRR system does not behave merely as a coupled oscillator network but instead functions as
a quantum interferometric platform~\cite{braakman2013coherent}, where hybridization is the measurable output of coherent phase competition between excitation and tunnelling channels~\cite{nazarov1993quantum,lee2018proximity}. Extended simulations for $\theta = 180^\circ$ and $\theta = 270^\circ$ are provided in the Supplementary Information (Section S5).This dual-phase mechanism allows programmable transitions between bright and dark states and supports applications in tunable interconnects, logic gates~\cite{chen2022qkd}, and non-Hermitian photonic devices~\cite{wu2024exact,hagness1998fdtd,gonzalez2024light,nagulu2021nonreciprocal}. Interactions involving a third spectral mode further enrich the response landscape, enabling multi-mode interference features such as selective filters and directional couplers~\cite{laucht2021roadmap}.

\section{Conclusion}
We have developed a quantum-interference-based framework to control photon tunneling in chiral IC-SRR systems by engineering both the structural orientation ($\phi$) and excitation phase ($\theta$). The coupling coefficient $\Delta_{AB}$ depends sensitively on these parameters, exhibiting a sign inversion beyond $\phi = 270^\circ$ that marks the transition from positive to negative coupling. Our model captures the magnitude, sign, and spatial decay of coupling, and predicts bright and dark mode formation through phase-tuned interference. Full-wave simulations confirm that $\theta$ can enhance or suppress hybridization, validating phase-control even under fixed geometry. This tunable coupling enables dynamic modulation of light–matter interactions in compact photonic platforms, with applications in phase-programmable isolators, dark-state filters, and reconfigurable resonator arrays. Our framework also supports chiral qudit logic, non-Hermitian effects, and coherent quantum photonic networks. The experimental observations across different chiral configurations further reinforce these conclusions and demonstrate the feasibility of implementing such interference-based control in practical device settings.

\section{acknowledgments}
\vspace{-1em}

The authors would like to thank Jayam Joshi for providing valuable insights and discussions during preparation of the manuscript. 
The work was supported by the Council of Science \& Technology, Uttar Pradesh (CSTUP), (Project Id: 2470, CST, U.P. sanction No: CST/D-1520 and Project Id: 4482, CST, U.P. sanction No: CST/D-7/8). 
B. Bhoi acknowledges support by the Science and Engineering Research Board (SERB) India—SRG/2023/001355.

\section{Statements and Declarations}
\textbf{Funding}: The authors declare grants or other support were received during the preparation of this manuscript.\\
\textbf{Competing Interests}: The authors declare that they have no competing interests. \\
\textbf{Author Contributions}: All authors contributed to the study's conception and design. R.S and B.B. led the work and wrote the manuscript with M. Ketkar and A. Srivastava. The other co-authors read, commented, and approved the final manuscript.

\section{Data Availability}

The data that support the findings of this study are available within the article.



\bibliographystyle{apsrev4-2}

\end{document}